\documentclass[a4paper]{article}%
\usepackage{amssymb}
\usepackage{amsmath}
\usepackage{graphicx}%
\usepackage{amsfonts}
\providecommand{\U}[1]{\protect\rule{.1in}{.1in}}

\begin{document}

\title{Eikonal Methods in AdS/CFT:\\Regge Theory and Multi--Reggeon Exchange}
\author{Lorenzo Cornalba\thanks{Email: \texttt{Lorenzo.Cornalba@mib.infn.it}}\medskip\\Universit\`{a} di Milano-Bicocca\\and INFN, sezione di Milano-Bicocca\\Piazza della Scienza 3, I--20126 Milano, Italy}
\date{}
\maketitle

\begin{abstract}
We analyze conformal field theory 4--point functions of the form
$A\sim\left\langle \mathcal{O}_{1}\left(  \mathbf{x}_{1}\right)
\mathcal{O}_{2}\left(  \mathbf{x}_{2}\right)  \mathcal{O}_{1}\left(
\mathbf{x}_{3}\right)  \mathcal{O}_{2}\left(  \mathbf{x}_{4}\right)
\right\rangle $, where the operators $\mathcal{O}_{i}$ are scalar primaries.
We show that, in the Lorentzian regime, the limit $\mathbf{x}_{1}%
\rightarrow\mathbf{x}_{3}$ is dominated by the exchange of
conformal partial waves of highest spin. When partial waves of
arbitrary spin contribute to
$A$, the behavior of the Lorentzian amplitude for $\mathbf{x}_{1}%
\rightarrow\mathbf{x}_{3}$ must be analyzed using complex--spin
techniques, leading to a generalized Regge theory for CFT's.
Whenever the CFT is dual to a string theory, the string
tree--level contribution $A_{\text{tree}}$ to the amplitude $A$
presents a Regge pole corresponding the a gravi--reggeon exchange.
In this case, we apply the impact parameter representation for CFT
amplitudes, previously developed, to analyze multiple reggeon
exchanges in the eikonal limit. As an example, we apply these
general techniques to $\mathcal{N}=4$ super--Yang--Mills theory in
$d=4$ in the limit of large 't Hooft coupling, including the
leading string corrections to pure graviton exchange.

\end{abstract}

\section{Introduction}

In this work, we wish to reconsider the theory of Regge poles
\cite{Analytic, Gribov} in the context of the AdS/CFT duality
\cite{Malda2}. Most of the discussion will be phrased uniquely in
terms of CFT data, and is therefore independent of the existence
of a gravity dual. On the other hand, whenever a string dual is
present, we expect that tree--level string interactions will be
dominated at high energy by a gravi--reggeon pole in AdS. In this
case, we shall show how to resum multi--reggeon interactions in
the eikonal limit, using an impact parameter representation for CFT
amplitudes developed in \cite{Paper1, Paper2, Paper3}. We take as
an example $\mathcal{N}=4$ $SU\left(  N\right)  $ supersymmetric
Yang--Mills theory in $d=4$ at large 't Hooft coupling, where the
gravi--reggeon pole corresponds dually to the pomeron trajectory \cite{PS1}.
We analyze the leading string corrections to pure graviton
exchange, following \cite{Paper3, PS1, PS2, PS3}.

Recall that, due to the presence of a time--like boundary, the AdS analogue of
the flat space scattering matrix computes correlators $\left\langle
\mathcal{O}_{1}\left(  \mathbf{x}_{1}\right)  \cdots\right\rangle $ in the
dual CFT. Moreover, the role played in flat space scattering by external
particle momenta is now played by the boundary points $\mathbf{x}_{i}$ in the dual
correlator. This interplay of AdS \textquotedblleft momenta\textquotedblright%
\ and CFT positions is at the heart of the simplicity of Regge theory for
CFT's in position space. In particular, we shall focus our attention on the
AdS analogue of $2\rightarrow2$ scattering in flat space, which computes the
conformal field theory correlator
\begin{equation}
A\sim\left\langle \mathcal{O}_{1}\left(  \mathbf{x}_{1}\right)  \mathcal{O}%
_{2}\left(  \mathbf{x}_{2}\right)  \mathcal{O}_{1}\left(  \mathbf{x}%
_{3}\right)  \mathcal{O}_{2}\left(  \mathbf{x}_{4}\right)  \right\rangle
_{\mathrm{CFT}}~,\label{i100}%
\end{equation}
where the operators $\mathcal{O}_{i}$ are scalar primaries. In analogy with a
flat space scattering process, we shall consider the operators $\mathcal{O}%
_{1}\left(  \mathbf{x}_{1}\right)  $ and $\mathcal{O}_{2}\left(
\mathbf{x}_{2}\right)  $ to be dual to incoming states, whereas $\mathcal{O}%
_{1}\left(  \mathbf{x}_{3}\right)  $ and $\mathcal{O}_{2}\left(
\mathbf{x}_{4}\right)  $ will be dual to outgoing ones. Accordingly, we shall
denote with S-- and T--channel the limits where $\mathbf{x}_{1}\rightarrow
\mathbf{x}_{2}$ and $\mathbf{x}_{1}\rightarrow\mathbf{x}_{3}$, respectively.

Let us first recall the situation in flat space scattering. At high energy,
interactions are dominated by the partial wave of highest spin $J$ exchanged
in the T--channel, with a characteristic growth of the amplitude given by
$s^{J}$. Whenever partial waves of all spins are present, the high energy
limit of the amplitude must be determined with the aid of a Sommerfeld--Watson
transform in the complex $J$--plane, which leads directly to Regge theory.
Tree--level string theory in flat space is a paradigmatic example of this
behavior, where massive higher spin interactions resum to an effective
gravi--reggeon pole with high energy behavior $s^{2+\frac{\alpha^{\prime}}%
{2}t}$. Moreover, in this case, higher loop string interactions can be
resummed in the eikonal limit \cite{ACV} at large transverse distances. This
is achieved by determining the phase shift using the impact parameter
representation of the leading tree level gravi--reggeon amplitude.

Consider now the situation in conformal field theory. In the Euclidean regime
of the CFT, the correlator (\ref{i100}) is determined, in the T--channel
$\mathbf{x}_{1}\rightarrow\mathbf{x}_{3}$, by the usual operator product
expansion and is therefore dominated by conformal partial waves of lowest
energy. The situation is therefore rather different from the analogous flat
space $2\rightarrow2$ S--matrix amplitude, where highest exchanged spin
dominates. On the other hand, if we consider the CFT amplitude in a specific
Lorentzian kinematical regime, the situation is analogous to a flat space
S--matrix \cite{Paper3}. More specifically, we shall consider the situation
where $\mathbf{x}_{4}$ is in the future of $\mathbf{x}_{1}$ and $\mathbf{x}%
_{3}$ is in the future of $\mathbf{x}_{2}$, with $\mathbf{x}_{1}%
,\mathbf{x}_{4}$ spacelike related to $\mathbf{x}_{2},\mathbf{x}_{3}$ (see
figure \ref{Fig1}). This
kinematics corresponds, dually, to a true $2\rightarrow2$ scattering process
in AdS and the limit $\mathbf{x}_{1}\rightarrow\mathbf{x}_{3}$ of a T--channel
partial wave of spin $J$ now behaves as $\left\vert \mathbf{x}_{3}%
-\mathbf{x}_{1}\right\vert ^{1-J}$. As in flat space, highest
spins now dominate. When the amplitude contains intermediate
conformal partial waves of all spins $J$, its limiting form for
$\mathbf{x}_{1}\rightarrow\mathbf{x}_{3}$ must be analyzed by
resumming all contributions in the complex $J$--plane. In this
paper, we show how to perform the relevant analytic continuation
in the complex $J$--plane and how to extract the leading behavior
of the Lorentzian amplitude in the limit
$\mathbf{x}_{1}\rightarrow\mathbf{x}_{3}$. In section
\ref{SecRegge}\ we shall describe, in particular, the behavior of the amplitude
$A$ when it is dominated by a leading Regge pole in the
$J$--plane. The pole is described by a trajectory $j\left(
\nu\right)  $, where $\nu$ is the AdS analogue of transverse
momentum transfer. More precisely, $\nu$ parameterizes the
spectral decomposition of the Laplacian operator on the relevant
transverse space, which turns out to be the hyperbolic space
$\mathrm{H}_{d-1}$, independently of the existence of a gravity
AdS dual \cite{Paper2}.

Finally, in section \ref{SecEikonal}, we consider theories which
admit a string dual description. The string tree level
contribution $A_{\text{tree}}$ to the full amplitude will then be
dominated, in the T--channel, by a leading Regge pole, which
corresponds to gravi--reggeon exchange in AdS. We then analyze
multi--reggeon exchanges in the limit $\mathbf{x}_{1}\rightarrow
\mathbf{x}_{3}$ using eikonal techniques. This is easily achieved
using the relevant impact parameter representation for the
tree--level amplitude $A_{\text{tree}}$, which was developed in
\cite{Paper2} and which determines the leading contribution to the
phase shift. Under the assumption that the phase shift is
determined only by the leading tree contribution for large impact
parameters, we are able to resum the contributions of
multi--reggeon exchanges, neglecting self--interactions, as in
flat space \cite{ACV}.

As an example of the general formalism, we conclude, in section
\ref{SecExample}, with the discussion of the gravi--reggeon
trajectory $j\left(  \nu,g\right)  $ dual to the pomeron
trajectory of $\mathcal{N}=4$ $SU\left(  N\right)  $
supersymmetric Yang--Mills theory in $d=4$. We analyze the leading
string corrections to the trajectory at large 't Hooft coupling,
quickly rederiving the results of \cite{PS1}. Our general
formalism then allows to resum multi--pomeron exchanges at large
't Hooft coupling in the large impact parameter limit. We will
extend this analysis to the weak 't Hooft coupling regime in
\cite{Paper5}.\medskip

NOTE ADDED: As this work was being completed, a paper appeared \cite{PS3}
which contains some overlapping results. The results in this paper have been
presented at the GGI Workshop "String and M theory approaches to particle
physics and cosmology", May 2007, at the 3rd RTN Workshop, Valencia 1--5
October 2007 and at Centro E. Fermi on September 5th, 2007.

\section{Basic Kinematics}

\subsection{General Setting\label{SecGeneral}}

\begin{figure}
[ptb]
\begin{center}
\includegraphics[
height=1.8688in,
width=4.0903in
]%
{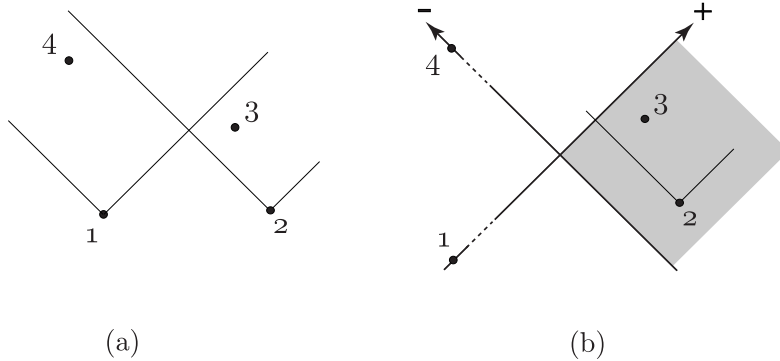}%
\caption{Basic Lorentzian kinematics for the correlator $A\left(
\mathbf{x}_{1},\mathbf{x}_{2},\mathbf{x}_{3},\mathbf{x}_{4}\right)  $. The
relevant light cones are shown in (a) and (b). The gray region in (b) with
$x^{+}>0$ and $x^{-}<0$ is mapped, under the conformal maps $\mathbf{\pi
,\bar{\pi}}$, to the future Milne cone $\mathrm{M}$.}%
\label{Fig4}%
\end{center}
\end{figure}

\begin{figure}
[ptb]
\begin{center}
\includegraphics[
height=1.9474in,
width=3.6379in
]%
{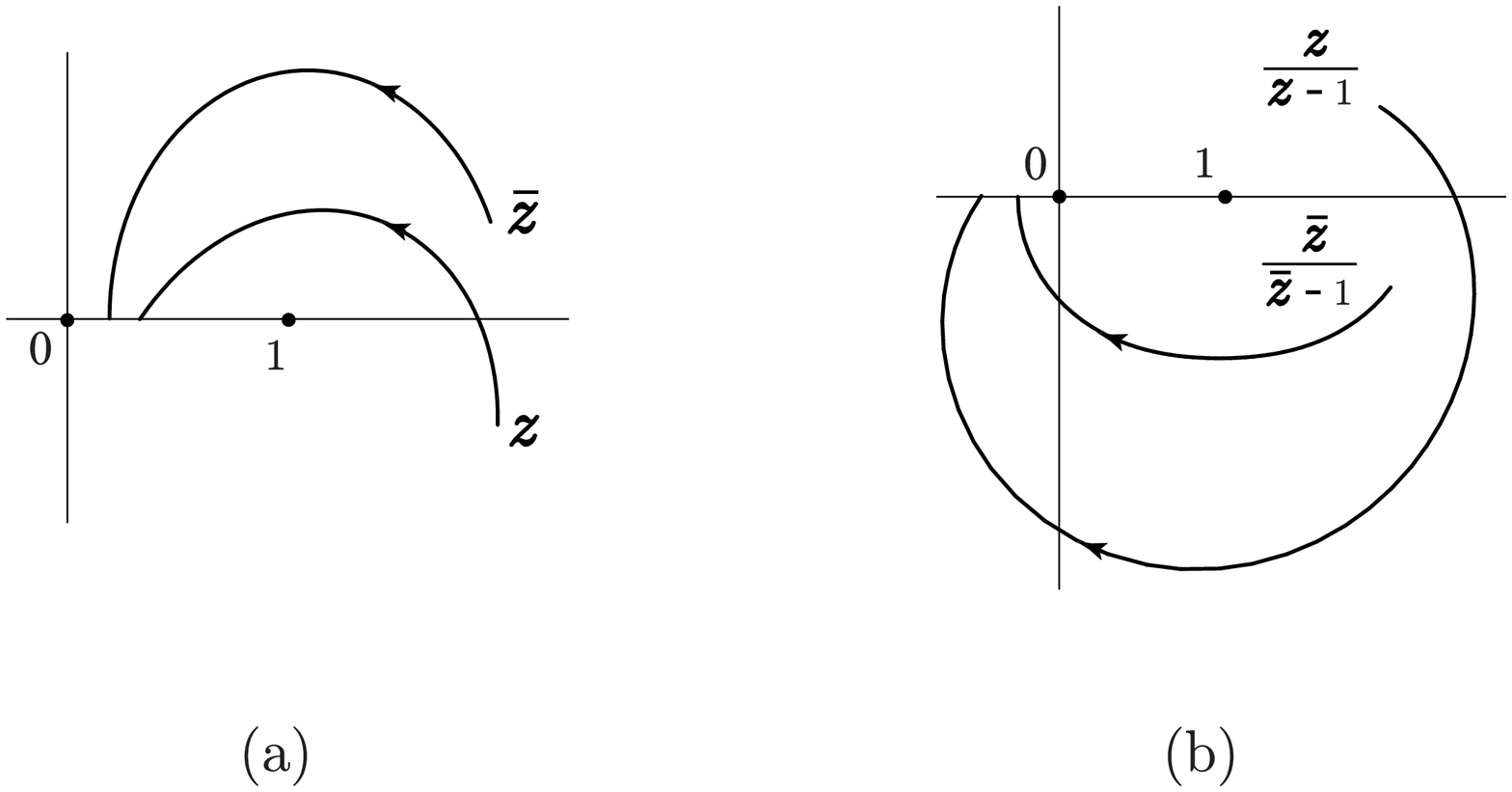}%
\caption{Relevant analytic continuation in $z,\bar{z}$ for the Lorentzian
amplitude $\hat{\mathcal{A}}$ starting from the Euclidean amplitude
$\mathcal{A}$ with $\bar{z}=z^{\star}$. In (b) we show the same paths under
the map $z\rightarrow z/(z-1)$ which exchanges the roles of particles $2$ and
$4$.}%
\label{Fig2}%
\end{center}
\end{figure}

In this first section, we shall review the results obtained in \cite{Paper1,
Paper2, Paper3} in a language which will be useful for the analysis of the
rest of the paper. Subject of our analysis will be a $d$--dimensional
conformal field theory CFT$_{d}$ defined on Minkowski space $\mathbb{M}^{d}$.
We parameterize points $\mathbf{x\in}\mathbb{M}^{d}$ with two lightcone
coordinates $x^{+},~x^{-}$ and with a point $x$ in the Euclidean transverse
space $\mathbb{E}^{d-2}$, and we choose metric%
\[
-dx^{+}dx^{-}+dx\cdot dx~.
\]
We will be interested in the analysis of the correlator%
\[
A\left(  \mathbf{x}_{1},\mathbf{x}_{2},\mathbf{x}_{3},\mathbf{x}_{4}\right)
=\left\langle \mathcal{O}_{1}\left(  \mathbf{x}_{1}\right)  \mathcal{O}%
_{2}\left(  \mathbf{x}_{2}\right)  \mathcal{O}_{1}\left(  \mathbf{x}%
_{3}\right)  \mathcal{O}_{2}\left(  \mathbf{x}_{4}\right)  \right\rangle
_{\mathrm{CFT}_{d}}~,
\]
where $\mathcal{O}_{1}$ and $\mathcal{O}_{2}$ are scalar primary operators of
dimension $\Delta_{1}$ and $\Delta_{2}$ respectively. As discussed in the
introduction, the above correlator is the CFT\ analogue of $2\rightarrow2$
scattering in a quantum field theory with a conventional scattering matrix.
Conformal invariance implies that $A$ can be written in terms of a reduced
amplitude $\mathcal{A}$ as%
\begin{equation}
A\left(  \mathbf{x}_{1},\mathbf{\cdots},\mathbf{x}_{4}\right)  =\frac
{1}{\left(  \mathbf{x}_{13}^{2}+i\epsilon\right)  ^{\Delta_{1}}\left(
\mathbf{x}_{24}^{2}+i\epsilon\right)  ^{\Delta_{2}}}\mathcal{A}\left(
z,\bar{z}\right)  ~,\label{eq100}%
\end{equation}
where $\mathbf{x}_{ij}=\mathbf{x}_{i}-\mathbf{x}_{j}$ and where we have
defined \cite{Osborn} the cross--ratios $z,\bar{z}$ by \footnote{Throughout
the paper, we shall consider barred and unbarred variables as independent,
with complex conjugation denoted by $\star$.}%
\[
z\bar{z}=\frac{\mathbf{x}_{13}^{2}\mathbf{x}_{24}^{2}}{\mathbf{x}_{12}%
^{2}\mathbf{x}_{34}^{2}}~,~\ \ \ \ \ \ \ \ \ \ \ \ \ \ \ \ \ \ \ \left(
1-z\right)  \left(  1-\bar{z}\right)  =\frac{\mathbf{x}_{14}^{2}%
\mathbf{x}_{23}^{2}}{\mathbf{x}_{12}^{2}\mathbf{x}_{34}^{2}}~.
\]
We are working in a Minkowskian setting, and therefore the causal relations
between the four points $\mathbf{x}_{i}$ is of primary importance. In
particular, we shall be interested in two basic configurations. Firstly, we
will consider the Euclidean configuration, where all points $\mathbf{x}_{i}$
are space--like related so that $\mathbf{x}_{ij}^{2}>0$. This first case is
directly related to the Euclidean version of the CFT, since the reduced
amplitude $\mathcal{A}$ is identical to the one of the Euclidean theory,
whenever $\bar{z}=z^{\star}$. We shall reserve from now on the name
$\mathcal{A}$ for the Euclidean amplitude. Of basic importance for our
discussion will be, on the other hand, configurations with a different causal
relation between the points $\mathbf{x}_{i}$, which really probe the
Lorentzian nature of the theory at hand and which are more closely related to
standard scattering theory. More precisely, we shall be interested in
configurations, called Lorentzian from now on, where%
\begin{align}
& \mathbf{x}_{4}\text{ in the future of }\mathbf{x}_{1}~,\nonumber\\
& \mathbf{x}_{3}\text{ in the future of }\mathbf{x}_{2}~,\label{eq200}\\
& \mathbf{x}_{12}^{2}~,\,\mathbf{x}_{13}^{2}~,\,\mathbf{x}_{24}^{2}%
~,\,\mathbf{x}_{34}^{2}~>0~,\nonumber
\end{align}
as shown in figure \ref{Fig4}a.%
As discussed extensively in \cite{Paper3}, in this regime the amplitude $A$
takes the same form (\ref{eq100}), with a reduced amplitude
\[
\hat{\mathcal{A}}\left(  z,\bar{z}\right)  =\left[  \mathcal{A}\left(
z,\bar{z}\right)  \right]  ^{\circlearrowleft}%
\]
related to the Euclidean amplitude $\mathcal{A}$ by analytic continuation.
More precisely, the relevant continuation $\circlearrowleft$ obtained by Wick
rotation is shown in figure \ref{Fig2}a and is given explicitly by%
\begin{align*}
& \bar{z}\text{ fixed~,}\\
& z\text{ counter--clockwise around }0,1~.
\end{align*}
For the purposes of this paper $z,\bar{z}$ are real and positive in
$\hat{\mathcal{A}}\left(  z,\bar{z}\right)  $, with an infinitesimal positive
imaginary part $+i\epsilon$.%

In what follows we shall restrict, without loss of generality due to conformal
invariance, to the following simpler configurations, as shown in figure
\ref{Fig4}b. We shall fix points $\mathbf{x}_{2},\mathbf{x}_{3}$ and send
$\mathbf{x}_{1},\mathbf{x}_{4}$ to infinity by setting $x_{1}=x_{4}=0$ and
$x_{1}^{-}=x_{4}^{+}=0$, and by considering the limit%
\[
A\left(  \mathbf{x}_{2}\mathbf{,x}_{3}\right)  =\left(  x_{3}^{+}\right)
^{\Delta_{1}}\left(  -x_{2}^{-}\right)  ^{\Delta_{2}}~\lim_{\substack{x_{1}%
^{+}\rightarrow-\infty\\x_{4}^{-}\rightarrow+\infty}}\left(  -x_{1}%
^{+}\right)  ^{\Delta_{1}}\left(  x_{4}^{-}\right)  ^{\Delta_{2}}A\left(
\mathbf{x}_{1},\mathbf{\cdots},\mathbf{x}_{4}\right)  ~.
\]
The Lorentzian conditions (\ref{eq200}) now simply imply that
\begin{align}
& x_{2}^{+},x_{3}^{+}>0~,\nonumber\\
& x_{2}^{-},x_{3}^{-}<0~,\label{eq300}\\
& \mathbf{x}_{3}\text{ in the future of }\mathbf{x}_{2}\mathbf{~.}\nonumber
\end{align}
The cross--ratios are then determined by%
\[
z\bar{z}=\frac{x_{3}^{-}x_{2}^{+}}{x_{3}^{+}x_{2}^{-}}%
~,~\ \ \ \ \ \ \ \ \ \ \ \left(  1-z\right)  \left(  1-\bar{z}\right)
=\frac{\mathbf{x}_{23}^{2}}{x_{3}^{+}x_{2}^{-}}~,
\]
and the amplitude is explicitly given by%
\[
A\left(  \mathbf{x}_{2}\mathbf{,x}_{3}\right)  =\left(  \frac{x_{3}^{+}%
}{-x_{3}^{-}}\right)  ^{\Delta_{1}}\left(  \frac{-x_{2}^{-}}{x_{2}^{+}%
}\right)  ^{\Delta_{2}}~\hat{\mathcal{A}}\left(  z,\bar{z}\right)  ~.
\]

\subsection{Transverse Space}

\begin{figure}
[ptb]
\begin{center}
\includegraphics[
height=1.8688in]%
{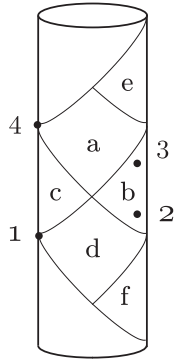}%
\caption{Boundary of AdS, with global time running vertically.
The CFT amplitude is originally defined on the Poincar\`{e} patch defined by the
regions a,b,c,d. On the other hand, it is more convenient to parameterize insertions of the operators
$\mathcal{O}_{1}$ and $\mathcal{O}_{2}$ at the points $3,2$ with coordinates $\mathbf{x},\mathbf{\bar x}$ 
relative to the left and right Poincar\`{e} patches. These are defined by
regions a,b,e and b,d,f, respectively, and are related to the original Poincar\`{e} patch
by the conformal transformations $\mathbf{\pi},\mathbf{\bar \pi}$.
}%
\label{Fig6}%
\end{center}
\end{figure}

In order to best understand the geometric nature of the correlator $A$, it is
often useful to use \cite{Paper1, Paper2, Paper3}, instead of the coordinates
$\mathbf{x}_{3},\mathbf{x}_{2}$, the related coordinates%
\[
\mathbf{x=\pi}\left(  \mathbf{x}_{3}\right)
~,~\ \ \ \ \ \ \ \ \ \ \ \ \ \ \ \ \ \ \mathbf{\bar{x}=\bar{\pi}}\left(
\mathbf{x}_{2}\right)  ~,
\]
where the two conformal transformations $\mathbf{\pi},\mathbf{\bar{\pi}%
}:\mathbb{M}^{d}\rightarrow\mathbb{M}^{d}$ are defined by%
\begin{align*}
& x^{+}=\frac{1}{x_{3}^{+}}~,\\
& x^{-}=-x_{3}^{-}+\frac{x_{3}\cdot x_{3}}{x_{3}^{+}}~,\\
& x=\frac{x_{3}}{x_{3}^{+}}~,
\end{align*}
and%
\begin{align*}
& \bar{x}^{+}=-\frac{1}{x_{2}^{-}}~,\\
& \bar{x}^{-}=x_{2}^{+}-\frac{x_{2}\cdot x_{2}}{x_{2}^{-}}~,\\
& \bar{x}=-\frac{x_{2}}{x_{2}^{-}}~.
\end{align*}
To understand the geometric meaning of the transformations $\mathbf{\pi
},\mathbf{\bar{\pi}}$, it is best to think of the CFT as being defined on the
boundary of global AdS \cite{Paper1, Paper2}, given by the light--cone
\begin{equation}
uv=\mathbf{w}^{2}=-w^{+}w^{-}+w\cdot w\label{eqsuca}%
\end{equation}
up to\ positive rescalings of the vector $\left(  u,v,\mathbf{w}\right)  $.
The CFT correlator $A$ is originally defined on the Poincar\`{e} patch
corresponding to the choice $u=1$. On the other hand, as shown in
\cite{Paper1, Paper2, Paper3} and in figure \ref{Fig6}, it is more convenient
geometrically to parameterize the four points of the amplitude $A$ using left
and right Poincar\`{e} patches, corresponding to insertions of the operators
$\mathcal{O}_{1}$ and $\mathcal{O}_{2}$ respectively. These patches are
defined by the choice $w^{+}=1$ and $w^{-}=-1$ respectively, inverting the
role of the light--cone coordinates $u,v$ and $w^{+},w^{-}$ in (\ref{eqsuca}).
For instance, the point $\mathbf{x}_{3}$ relative to operator $\mathcal{O}%
_{1}$ corresponds to the vector $\left(  u,v,\mathbf{w}\right)  =\left(
1,\mathbf{x}_{3}^{2},\mathbf{x}_{3}\right)  $. Rescaling by $x_{3}^{+}$ and
inverting the role of the light--cone coordinates we obtain the map
$\mathbf{\pi}$. Similarly, the map $\mathbf{\bar{\pi}}$ is obtained by
rescaling $\left(  1,\mathbf{x}_{2}^{2},\mathbf{x}_{2}\right)  $ by
$-x_{2}^{-}$.

Note that%
\[
\mathbf{x}^{2}=\frac{x_{3}^{-}}{x_{3}^{+}}%
~,~\ \ \ \ \ \ \ \ \ \ \ \ \mathbf{\bar{x}}^{2}=\frac{x_{2}^{+}}{x_{2}^{-}}~,
\]
and that the regions $x_{3}^{+},-x_{3}^{-}>0$ and $x_{2}^{+},-x_{2}^{-}>0$ are
mapped, by the transformations $\mathbf{\pi,\bar{\pi}}$, into the future Milne
cone $\mathrm{M}\subset\mathbb{M}^{d}$ of Minkowski space. Therefore, the
Lorentzian conditions (\ref{eq300}) now imply that
\begin{align*}
& \mathbf{x,\bar{x}}\in\mathrm{M}~,\\
& -\mathbf{x/x}^{2}\text{ in the future of }\mathbf{\bar{x}~.}%
\end{align*}
The cross--ratios are given by
\[
z\bar{z}=\mathbf{x}^{2}\mathbf{\bar{x}}^{2}%
~,~\ \ \ \ \ \ \ \ \ \ \ \ \ \ \ \ \ z+\bar{z}=-2\mathbf{x\cdot\bar{x}},
\]
and the amplitude $A$ now reads%
\begin{equation}
A\left(  \mathbf{x},\mathbf{\bar{x}}\right)  =~\frac{1}{\left(  -\mathbf{x}%
^{2}\right)  ^{\Delta_{1}}\left(  -\mathbf{\bar{x}}^{2}\right)  ^{\Delta_{2}}%
}\hat{\mathcal{A}}\left(  z,\bar{z}\right)  ~.\label{eq2000}%
\end{equation}
In what follows, it will be often convenient to denote the cross--ratios as%
\begin{align*}
z  & =\frac{\sigma}{4}\,e^{\rho}~,\\
\bar{z}  & =\frac{\sigma}{4}\,e^{-\rho}~.
\end{align*}
In this notation, $\sigma=4\left\vert \mathbf{x}\right\vert \left\vert
\mathbf{\bar{x}}\right\vert $ and $\rho$ is the geodesic distance between the
points%
\[
\frac{\mathbf{x}}{\left\vert \mathbf{x}\right\vert },\frac{\mathbf{\bar{x}}%
}{\left\vert \mathbf{\bar{x}}\right\vert }\in\mathrm{H}_{d-1}~,
\]
where $\mathrm{H}_{d-1}$ $\subset\mathrm{M}\subset\mathbb{M}^{d}$ represents
the transverse hyperboloid of points $\mathbf{w}\in\mathrm{M}$ with
$\mathbf{w}^{2}=-1$.

In some situations, and again without loss of generality due to conformal
invariance, we shall find it convenient to restrict the kinematics so that
$\mathbf{x}_{2},\mathbf{x}_{3}$, and therefore $\mathbf{x,\bar{x}}$, have
vanishing transverse components $x_{2}=x_{3}=x=\bar{x}=0$. Then we have that%
\[
z=\frac{x_{3}^{-}}{x_{2}^{-}}=x^{-}\bar{x}^{+}%
~,~\ \ \ \ \ \ \ \ \ \ \ \ \ \ \ \ \ \bar{z}=\frac{x_{2}^{+}}{x_{3}^{+}}%
=x^{+}\bar{x}^{-}~,
\]
and%
\begin{align*}
A  & =~\left(  \frac{x_{3}^{+}}{-x_{3}^{-}}\right)  ^{\Delta_{1}}\left(
\frac{-x_{2}^{-}}{x_{2}^{+}}\right)  ^{\Delta_{2}}~\hat{\mathcal{A}}\left(
\frac{x_{3}^{-}}{x_{2}^{-}},\frac{x_{2}^{+}}{x_{3}^{+}}\right) \\
& =\frac{1}{\left(  x^{+}x^{-}\right)  ^{\Delta_{1}}\left(  \bar{x}^{+}\bar
{x}^{-}\right)  ^{\Delta_{2}}}~~\hat{\mathcal{A}}\left(  x^{-}\bar{x}%
^{+},x^{+}\bar{x}^{-}\right)  ~.
\end{align*}

\subsection{The Impact Parameter Representation}

To conclude this introductory section, we briefly recall the results of
\cite{Paper2, Paper3} on the impact parameter representation of $A$ in the
Lorentzian regime. This is obtained by considering the Fourier transform $B$
of the amplitude \footnote{Throughout this paper we shall use the notation
$\left(  -\right)  ^{\alpha}\equiv e^{i\pi\alpha}$.}
\begin{equation}
A\left(  \mathbf{x},\mathbf{\bar{x}}\right)  =\left(  -\right)  ^{-\Delta_{1}%
}\left(  -\right)  ^{-\Delta_{2}}\int_{\mathrm{M}}d\mathbf{y}d\mathbf{\bar{y}%
}~e^{-2i\mathbf{x\cdot y-}2i\mathbf{\bar{x}\cdot\bar{y}}}~B\left(
\mathbf{y},\mathbf{\bar{y}}\right)  ~,\label{eq3000}%
\end{equation}
which has support uniquely in the future Milne cone $\mathrm{M}$. The
amplitude $B$ has a representation similar to (\ref{eq2000})
\begin{equation}
B\left(  \mathbf{y},\mathbf{\bar{y}}\right)  =\frac{\mathcal{N}_{\Delta_{1}}%
}{\left(  -\mathbf{y}^{2}\right)  ^{\frac{d}{2}-\Delta_{1}}}~\frac
{\mathcal{N}_{\Delta_{2}}}{\left(  -\mathbf{\bar{y}}^{2}\right)  ^{\frac{d}%
{2}-\Delta_{2}}}~\mathcal{B}\left(  h,\bar{h}\right)  ~,\label{eq3100}%
\end{equation}
where
\[
\mathcal{N}_{\Delta}=\frac{2\pi~\pi^{-\frac{d}{2}}}{\Gamma\left(
\Delta\right)  \Gamma\left(  1+\Delta-\frac{d}{2}\right)  }%
\]
and where the reduced amplitude $\mathcal{B}$ depends on $\mathbf{y}%
,\mathbf{\bar{y}}$ through the effective cross--ratios $h,\bar{h}$ defined by%
\[
h^{2}\bar{h}^{2}=\mathbf{y}^{2}\mathbf{\bar{y}}^{2}%
~,~\ \ \ \ \ \ \ \ \ \ \ \ \ \ \ \ \ h^{2}+\bar{h}^{2}=-2\mathbf{y\cdot\bar
{y}}.
\]
We shall also use, instead of $h,\bar{h}$, the equivalent ratios $s,r$,
defined by%
\begin{align*}
h^{2}  & =\frac{s}{4}\,e^{r}~,\\
\bar{h}^{2}  & =\frac{s}{4}\,e^{-r}~,
\end{align*}
where again $r$ is the geodesic distance between the points $\mathbf{y/}%
\left\vert \mathbf{y}\right\vert $ and $\mathbf{\bar{y}/}\left\vert
\mathbf{\bar{y}}\right\vert $ in $\mathrm{H}_{d-1}$. Recall \cite{Paper3} that
it is often convenient to express $\mathcal{B}$ as%
\begin{equation}
\mathcal{B}\left(  h,\bar{h}\right)  =e^{-2\pi i~\Gamma\left(  h,\bar
{h}\right)  }~,\label{eq20000}%
\end{equation}
where $\Gamma\left(  h,\bar{h}\right)  $ plays the dual role of phase shift
$-\pi\Gamma$ in the AdS dual of the CFT under consideration, and of anomalous
dimension $2\Gamma$ of the composites $\mathcal{O}_{1}\partial\cdots
\partial\mathcal{O}_{2}$ at large values of the energy $h+\bar{h}$ and spin
$h-\bar{h}$.

\section{Regge Theory in CFT\label{SecRegge}}

In this section we shall use the results of \cite{Paper2} to develop a Regge
theory suitable for conformal field theories. From now on, we shall continue
the discussion assuming $d=4$. This is done uniquely for notational
simplicity, since most formul\ae \ generalize to general $d$, as shown in
Appendix \ref{AppBis}.

\subsection{T--Channel Partial Waves}

We start by considering the contribution to the amplitude $A$ coming from the
exchange of a primary operator of \footnote{We shall always write dimensions
$\Delta$ of operators as $\frac{d}{2}+E$, so that the corresponding scalar
mass $\Delta\left(  \Delta-d\right)  $ is real either for $E$ purely real, or
for $E=i\nu$ purely imaginary. This convention is useful also when considering
high energy interactions. As shown in \cite{Paper1, Paper2, Paper3}, eikonal
interactions mediated by a particle of dimension $\Delta$ are effectively
described by a scalar particle in the transverse $d-2$ dimensions. The
conformal dimension of this effective particle is $\Delta-1$, and therefore
the energy $E$ is invariant.}
\begin{align*}
& \text{energy }2+E~,\\
& \text{spin }J~,
\end{align*}
in the T--channel $\mathbf{x}_{1}\rightarrow\mathbf{x}_{3}$, $\mathbf{x}%
_{2}\rightarrow\mathbf{x}_{4}$. This corresponds to the limit $z,\bar
{z}\rightarrow0$ or to
\[
\sigma\rightarrow0~.
\]
We shall denote the corresponding contribution to the reduced Euclidean
amplitude $\mathcal{A}$ by%
\[
\mathcal{T}_{E,J}\,\left(  z,\bar{z}\right)  ~.
\]
The expression for $\mathcal{T}_{E,J}$ can be written, in $d=4$, in terms of
the chiral expression%
\begin{equation}
\mathcal{T}_{k}\left(  z\right)  =\left(  -z\right)  ^{k}F\left(
k,k,2k,z\right)  ~,\label{eq2300}%
\end{equation}
with $F$ the hypergeometric function $_{2}F_{1}$. The partial waves
$\mathcal{T}_{E,J}$ are then explicitly given by \cite{Osborn}%
\[
\mathcal{T}_{E,J}\left(  z,\bar{z}\right)  =-\frac{z\bar{z}}{z-\bar{z}}\left[
\mathcal{T}_{k}\left(  z\right)  \mathcal{T}_{\bar{k}-1}\left(  \bar
{z}\right)  -\mathcal{T}_{k}\left(  \bar{z}\right)  \mathcal{T}_{\bar{k}%
-1}\left(  z\right)  \right]  ~,
\]
where
\begin{align*}
k+\bar{k}  & =2+E~,\\
k-\bar{k}  & =J~.
\end{align*}

The limit $\sigma\rightarrow0$ of $\mathcal{T}_{E,J}$ is determined by the
usual operator product expansion, and we have chosen the normalization of
$\mathcal{T}_{E,J}$ so that, for integer spin $J$, we have%
\begin{align*}
\mathcal{T}_{E,J}  & \sim\sum_{m=0}^{J}\left(  -z\right)  ^{k-m}\left(
-\bar{z}\right)  ^{\bar{k}+m}\\
& \sim\left(  -\right)  ^{J}\left(  \frac{\sigma}{4}\right)  ^{2+E}%
~\frac{\sinh\left(  \left(  J+1\right)  \rho\right)  }{\sinh\left(
\rho\right)  }~.
\end{align*}
Clearly, the leading behavior of the amplitude $A$ for $\sigma\rightarrow0$ is
determined by the operators of lowest energy. As shown in \cite{Paper2}, the
$\sigma\rightarrow0$ limit of the Lorentzian partial wave $\mathcal{\hat{T}%
}_{E,J}$ is, on the other hand, rather different. It is given explicitly by
\footnote{The expressions which follow are valid for general complex $J$. The
leading behavior $\mathcal{\hat{T}}_{E,J}\sim\sigma^{1-J}$ is valid for
$\operatorname{Re}J>-1$, and always dominates $\sigma^{2+E}$ for
$\operatorname{Re}E>0$. For $\operatorname{Re}J<-1$ we must replace below
$J\rightarrow-2-J$, since $\mathcal{T}_{E,J}=-\mathcal{T}_{E,-2-J}$.}%
\[
\mathcal{\hat{T}}_{E,J}\sim2\pi i~\left(  -\right)  ^{J}~\left(  \frac{\sigma
}{4}\right)  ^{1-J}~~\frac{\Pi_{E}\left(  \rho\right)  }{\tau\left(
J+E\right)  }~,
\]
where the coefficient $\tau\left(  x\right)  $ is given by%
\[
\tau\left(  x\right)  =-\frac{1}{2\pi}\,\frac{~\Gamma\left(  1+\frac{x}%
{2}\right)  ^{4}}{\Gamma\left(  2+x\right)  \Gamma\left(  1+x\right)  }~,
\]
and where%
\begin{equation}
\Pi_{E}\left(  \rho\right)  =\frac{1}{4\pi}\frac{e^{-E\rho}}{\sinh\rho
}\label{eq400}%
\end{equation}
is the scalar propagator in the transverse hyperbolic space $\mathrm{H}_{3}$
of dimension $1+E$, normalized as%
\begin{equation}
\left[  \square_{\mathrm{H}_{3}}-E^{2}+1\right]  ~\Pi_{E}=-\delta
_{\mathrm{H}_{3}}~.\label{eq2100}%
\end{equation}
We see that the behavior for $\sigma\rightarrow0$ is now controlled by the
operators of highest spin which are exchanged in the T--channel. The
$\sigma\rightarrow0$ limit of the Lorentzian amplitude is then much richer in
structure then the corresponding limit of the Euclidean amplitude. In fact, at
a fixed energy, we can have only a finite number of partial waves due to the
unitarity bounds of the conformal field theory and to the discreteness of the
spin variable. On the other hand, at fixed spin, we have a continuous of
possible energy contributions.

\subsection{Regge Poles in CFT}

The situation in the Lorentzian regime is rather similar to the usual partial
wave decomposition in flat space. In that case, one decompose amplitudes as
$\sum_{J}\mathcal{T}_{J}\left(  -s/t\right)  \,a_{J}\left(  t\right)  $, where
$s,t$ are the Mandelstam invariants and where the partial waves $\mathcal{T}%
_{J}$ are now polynomials of degree $J$. In the high energy limit
$s\rightarrow\infty\,$\ with $t$ fixed, a single partial wave behaves as
$s^{J}$ and thus highest spin waves dominate. On the other hand, if there are
contributions from partial waves of all spins, one must resum all
contributions in order to analyze the high $s$ behavior of the amplitude. In
the flat space case, this resummation is well understood in terms of the
Sommerfeld--Watson transform, and leads to the theory of Regge poles
\cite{Analytic, Gribov}. We shall see that the extension of these usual
techniques to conformal field theories is rather straight--forward.

Before starting the discussion, we need some basic results regarding harmonic
analysis of functions on the transverse space $\mathrm{H}_{3}$. More
precisely, we shall consider functions $G\left(  \mathbf{w},\mathbf{w}%
^{\prime}\right)  $, with $\mathbf{w},\mathbf{w}^{\prime}\mathbf{\in
}\mathrm{H}_{3}$, which depend only on the geodesic distance $\rho$ between
$\mathbf{w},\mathbf{w}^{\prime}$, given by $\cosh\rho=-\mathbf{w}%
\cdot\mathbf{w}^{\prime}$. We shall call those functions radial for short,
with an example given by the propagator $\Pi_{E}\left(  \mathbf{w}%
,\mathbf{w}^{\prime}\right)  $ in (\ref{eq400}). In order to analyze radial
functions, we first define $\Omega_{E}\left(  \mathbf{w},\mathbf{w}^{\prime
}\right)  $ as%
\begin{align}
\Omega_{E}  & =\frac{E}{2\pi}\left(  \Pi_{E}-\Pi_{-E}\right)  ~,\label{eq800}%
\\
& =-\frac{E}{4\pi^{2}}\frac{\sinh E\rho}{\sinh\rho}~,\nonumber
\end{align}
which satisfies the homogeneous equation
\begin{equation}
\left[  \square_{\mathrm{H}_{3}}-E^{2}+1\right]  ~\Omega_{E}=0~,\label{eq2200}%
\end{equation}
and which is even in $E$, since $\Omega_{E}=\Omega_{-E}$. It should be clear
to the reader that the functions $\Omega_{i\nu}$ with $E=i\nu$ pure imaginary
are the analogues, in $\mathrm{H}_{3}$, of the usual radial Bessel functions
which give a Fourier basis for radial functions in $\mathbb{E}^{3}$. We then
expect that a general radial function $G\left(  \mathbf{w},\mathbf{w}^{\prime
}\right)  $ can be expressed as%
\begin{equation}
G\left(  \mathbf{w},\mathbf{w}^{\prime}\right)  =\int d\nu~g\left(
\nu\right)  ~\Omega_{i\nu}\left(  \mathbf{w},\mathbf{w}^{\prime}\right)
~,\label{eq1000}%
\end{equation}
with $g\left(  \nu\right)  =g\left(  -\nu\right)  $. For example, we have that%
\begin{equation}
\Pi_{E}=\int d\nu~\frac{\Omega_{i\nu}}{\nu^{2}+E^{2}}~~.\label{eq600}%
\end{equation}
This can be shown by first using the definition of $\Omega_{E}$ to rewrite the
right hand side of the above equation as%
\[
-2\int\frac{d\nu}{2\pi i}~\frac{\nu}{\nu^{2}+E^{2}}\Pi_{i\nu}~,
\]
and then by closing the $\nu$--contour in the $\operatorname{Im}\nu\leq0$
region, noting that $\Pi_{i\nu}$ is free of singularities and decays
exponentially at infinity. The transform (\ref{eq1000}) can be inverted by
first noting that%
\begin{equation}
2\int_{\mathrm{H}_{3}}d\mathbf{w}^{\prime}~\Omega_{i\nu}\left(  \mathbf{w}%
,\mathbf{w}^{\prime}\right)  \,\Omega_{i\bar{\nu}}\left(  \mathbf{w}^{\prime
},\mathbf{w}^{\prime\prime}\right)  =\left[  \delta\left(  \nu-\bar{\nu
}\right)  +\delta\left(  \nu+\bar{\nu}\right)  \right]  ~\Omega_{i\nu}\left(
\mathbf{w},\mathbf{w}^{\prime\prime}\right)  ~.\label{eq700}%
\end{equation}
The normalization in the above expression can be fixed by integrating in $\nu$
and by using the fact that%
\[
\delta_{\mathrm{H}_{3}}\left(  \mathbf{w},\mathbf{w}^{\prime}\right)  =\int
d\nu~\Omega_{i\nu}\left(  \mathbf{w},\mathbf{w}^{\prime}\right)  ~,
\]
which immediately follows from (\ref{eq600}), (\ref{eq2100}) and
(\ref{eq2200}). We then have that%
\[
g\left(  \nu\right)  \Omega_{i\nu}\left(  \mathbf{w},\mathbf{w}^{\prime\prime
}\right)  =\int_{_{\mathrm{H}_{3}}}d\mathbf{w}^{\prime}~\Omega_{i\nu}\left(
\mathbf{w},\mathbf{w}^{\prime}\right)  ~G\left(  \mathbf{w}^{\prime
},\mathbf{w}^{\prime\prime}\right)
\]
and, choosing $\mathbf{w}=\mathbf{w}^{\prime\prime}$, that
\begin{align*}
g\left(  \nu\right)   & =\frac{1}{\Omega_{i\nu}\left(  \mathbf{w}%
,\mathbf{w}\right)  }\int_{_{\mathrm{H}_{3}}}d\mathbf{w}^{\prime}~\Omega
_{i\nu}\left(  \mathbf{w},\mathbf{w}^{\prime}\right)  ~G\left(  \mathbf{w}%
^{\prime},\mathbf{w}\right)  ~,\\
& =\frac{4\pi}{\nu}\int_{0}^{\infty}d\rho~\sin\nu\rho~\sinh\rho~G\left(
\rho\right)  ~.
\end{align*}
To conclude this preliminary discussion, we note that (\ref{eq700}) implies
that convolution $\int_{\mathrm{H}_{3}}d\mathbf{w}^{\prime}~G\left(
\mathbf{w},\mathbf{w}^{\prime}\right)  F\left(  \mathbf{w}^{\prime}%
,\mathbf{w}^{\prime\prime}\right)  $ in $\mathrm{H}_{3}$ goes to the product
$g\left(  \nu\right)  f\left(  \nu\right)  $ in Fourier space, as usual.

We may now start the discussion of Regge theory by recalling that, in the
previous section, we had shown the limit, in the Lorentzian regime and for
$\sigma\rightarrow0$, of%
\[
\tau\left(  J+E\right)  ~\mathcal{T}_{E,J}~\ \ \ \ \ \ \ \rightarrow
\ \ ~\ \ ~~~2\pi i~\left(  -\right)  ^{J}\left(  \frac{\sigma}{4}\right)
^{1-J}~~\Pi_{E}\left(  \rho\right)  ~.
\]
In analogy with (\ref{eq800}), we then introduce the composite partial waves%
\[
\mathcal{G}_{E,J}=\frac{E}{2\pi}~\left[  \tau\left(  J+E\right)
~\mathcal{T}_{E,J}-\tau\left(  J-E\right)  ~\mathcal{T}_{-E,J}\right]  ~,
\]
so that we have the correspondence
\[
\mathcal{G}_{E,J}~\ \ \ \ \ \ \ \rightarrow\ \ ~\ \ ~~2\pi i~\left(  -\right)
^{J}\left(  \frac{\sigma}{4}\right)  ^{1-J}~~\Omega_{E}\left(  \rho\right)  ~.
\]
Equation (\ref{eq1000}) then suggests to consider the following generalized
integrals%
\begin{equation}
\int d\nu~g_{J}\left(  \nu\right)  ~\mathcal{G}_{i\nu,J}~,\label{eq1300}%
\end{equation}
with $g_{J}\left(  \nu\right)  =g_{J}\left(  -\nu\right)  $ and with
$\nu\,g_{J}\left(  \nu\right)  $ square--integrable, in order to have well
defined transforms. To compute the integral above, we first write it as%
\[
-2\int\frac{d\nu}{2\pi i}~~g_{J}\left(  \nu\right)  ~\nu~\tau\left(
J+i\nu\right)  ~\mathcal{T}_{i\nu,J}%
\]
and we then close the contour in the region $\operatorname{Im}\nu\leq0$ to get%
\begin{equation}
-2i\sum_{E}\operatorname*{Res}g_{J}\left(  -iE\right)  \cdot E~\tau\left(
J+E\right)  ~\mathcal{T}_{E,J}~+\cdots~.\label{eq1100}%
\end{equation}
We have only shown the contributions coming from the poles of $g_{J}\left(
\nu\right)  $, with the omitted $\cdots$ indicating contributions from the
poles of the partial waves $\mathcal{T}_{i\nu,J}$, since $\nu~\tau\left(
J+i\nu\right)  $ is regular in the region of interest. These extra
contributions are given by partial waves $\mathcal{T}_{J+1,J^{\prime}}~$of
lower spin $J^{\prime}=J-2,~J-4,\cdots$ and fixed energy $J+1$, whose precise
form will not be fundamental for the main discussion of this paper. We shall
therefore describe them in detail in a separate section \ref{Sect100} in
equation (\ref{eq1200}), mostly for completeness. In any case, since the
omitted contribution is built explicitly out of partial waves of of lower spin
$J^{\prime}<J$, the leading contribution to the Lorentzian amplitude in the
$\sigma\rightarrow0$ limit coming from (\ref{eq1300}) is given by%
\begin{equation}
\int d\nu~g_{J}\left(  \nu\right)  ~\hat{\mathcal{G}}_{i\nu,J}\ ~\ \ \underset
{\sigma\rightarrow0}{\rightarrow}~\ \ ~2\pi i~\left(  -\right)  ^{J}\left(
\frac{\sigma}{4}\right)  ^{1-J}~G_{J}\left(  \rho\right)  ~,\label{eq1400}%
\end{equation}
with%
\[
G_{J}\left(  \rho\right)  =\int d\nu~g_{J}\left(  \nu\right)  ~\Omega_{i\nu
}\left(  \rho\right)  ~.
\]
The full amplitude $\mathcal{A}$ will have then a decomposition in terms of
$\mathrm{T}$--channel conformal partial waves given by%
\[
\mathcal{A}=\sum_{J\geq0}\int d\nu~g_{J}\left(  \nu\right)  ~\mathcal{G}%
_{i\nu,J}~.
\]

\begin{figure}
[ptb]
\begin{center}
\includegraphics[
height=2.1745in,
width=2.7477in
]%
{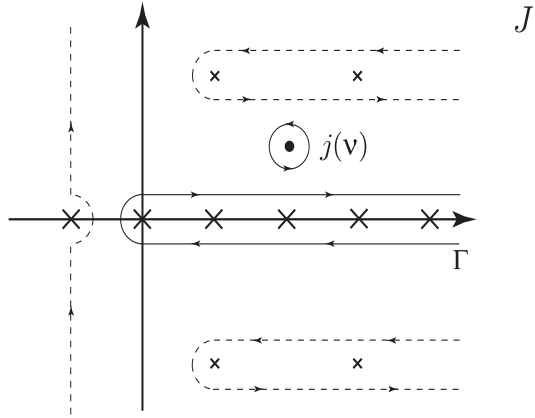}%
\caption{The path $\Gamma$ in the complex $J$--plane, which gives the usual
sum over integral spins, is deformed to analyze the $\sigma\rightarrow0$ limit
of the Lorentzian amplitude. Contributions come from Regge poles and cuts (a
pole is shown at $J=j\left(  \nu\right)  $), together with subdominant shadow
contributions coming from the dashed contours. This includes the usual path at
$\operatorname{Re}J=-1$ together with the contributions of the spurious poles
at $J=\pm i\nu+m$ with $m$ a positive odd integer, which have an effective
spin $-1\pm i\nu$.}%
\label{Fig1}%
\end{center}
\end{figure}

In order to understand the behavior of the Lorentzian amplitude $\hat
{\mathcal{A}}$ in the limit $\sigma\rightarrow0$, we clearly cannot use
(\ref{eq1400}) directly whenever the sum over $J$ in the above decomposition
is unbounded. We must therefore use, as in flat space, a Sommerfeld--Watson
resummation, which starts by separating the contributions of even and odd
spins as $\mathcal{A}=\mathcal{A}_{+}+\mathcal{A}_{-}$, with
\[
\mathcal{A}_{\pm}=\frac{1}{2}\sum_{J\geq0}\int d\nu~g_{J}\left(  \nu\right)
~\left[  \mathcal{G}_{i\nu,J}\left(  z,\bar{z}\right)  \pm\mathcal{G}_{i\nu
,J}\left(  \frac{z}{z-1},\frac{\bar{z}}{\bar{z}-1}\right)  \right]  ~.
\]
Note that the transformation $z\rightarrow z/\left(  z-1\right)  $ and
$\bar{z}\rightarrow\bar{z}/\left(  \bar{z}-1\right)  $ interchanges the roles
of the operators at $\mathbf{x}_{2}$ and $\mathbf{x}_{4}$, and in flat space
corresponds to the exchange of the roles of the Mandelstam variables $s,u$.
Moreover it is easy to show that, for $J$ integral,%
\[
\mathcal{T}_{E,J}\left(  \frac{z}{z-1},\frac{\bar{z}}{\bar{z}-1}\right)
=\left(  -\right)  ^{J}~\mathcal{T}_{E,J}\left(  z,\bar{z}\right)
~,~\ \ \ \ \ \ \ \ \ \ \left(  J\in\mathbb{Z}\right)
\]
with a similar equation for $\mathcal{G}_{E,J}$. Extending $g_{J}\left(
\nu\right)  $ in the complex $J$--plane to $g_{\pm}\left(  \nu,J\right)  $, we
may write an explicit Sommerfeld--Watson representation for $\mathcal{A}_{\pm
}$
\begin{equation}
\mathcal{A}_{\pm}=\frac{i}{4}\int d\nu\frac{dJ}{\sin\pi J}~g_{\pm}\left(
\nu,J\right)  ~\left[  \mathcal{G}_{i\nu,J}\left(  \frac{z}{z-1},\frac{\bar
{z}}{\bar{z}-1}\right)  \pm\mathcal{G}_{i\nu,J}\left(  z,\bar{z}\right)
\right]  ~,\label{eq3400}%
\end{equation}
where the $J$--contour of integration picks up the zeros of $\sin\pi J$ for
$J\geq0$. As shown in figure \ref{Fig1}, we shall deform the $J$ integration
contour, assuming that the functions $g_{\pm}\left(  \nu,J\right)  $ have a
leading pole%
\[
g_{\pm}\left(  \nu,J\right)  \sim\frac{r_{\pm}\left(  \nu\right)  }{J-j\left(
\nu\right)  }%
\]
along the Regge trajectory $J=j\left(  \nu\right)  $, with residue $r_{\pm
}\left(  \nu\right)  $.
The contribution to the amplitude $\mathcal{A}$ from this pole is then given
by
\[
\mathcal{A}\simeq-\frac{\pi}{2}\int d\nu~\frac{r_{\pm}\left(  \nu\right)
}{\sin\pi j\left(  \nu\right)  }~\left[  \mathcal{G}_{i\nu,j\left(
\nu\right)  }\left(  \frac{z}{z-1},\frac{\bar{z}}{\bar{z}-1}\right)
\pm\mathcal{G}_{i\nu,j\left(  \nu\right)  }\left(  z,\bar{z}\right)  \right]
~.
\]
We are now interested in the $\sigma\rightarrow0$ limit in the Lorentzian
regime. In this regime we have that%
\begin{align*}
\mathcal{G}_{i\nu,j\left(  \nu\right)  }\left(  z,\bar{z}\right)   & \sim2\pi
i~\left(  -\right)  ^{j\left(  \nu\right)  }\left(  \frac{\sigma}{4}\right)
^{1-j\left(  \nu\right)  }~~\Omega_{i\nu}\left(  \rho\right)  ~,\\
\mathcal{G}_{i\nu,j\left(  \nu\right)  }\left(  \frac{z}{z-1},\frac{\bar{z}%
}{\bar{z}-1}\right)   & \sim2\pi i~\left(  \frac{\sigma}{4}\right)
^{1-j\left(  \nu\right)  }\Omega_{i\nu}\left(  \rho\right)  ~,
\end{align*}
where the relevant analytic continuations are shown in figure \ref{Fig2}a,b.
We then obtain%
\begin{equation}
\hat{\mathcal{A}}\simeq2\pi i\int d\nu~\left(  -\right)  ^{j\left(
\nu\right)  }~\alpha\left(  \nu\right)  ~\sigma^{1-j\left(  \nu\right)
}~\Omega_{i\nu}\left(  \rho\right)  ~,\label{eq3200}%
\end{equation}
where $\alpha\left(  \nu\right)  $ is given, for poles of positive and
negative signature, by%
\begin{align*}
\alpha\left(  \nu\right)   & =\left(  -\right)  ^{\frac{2-j\left(  \nu\right)
}{2}}~\frac{\pi\,}{2}\,4^{j\left(  \nu\right)  -1}~\frac{r_{+}\left(
\nu\right)  }{\sin\frac{\pi j\left(  \nu\right)  }{2}}~,\\
\alpha\left(  \nu\right)   & =\left(  -\right)  ^{\frac{1-j\left(  \nu\right)
}{2}}~\frac{\pi\,}{2}\,4^{j\left(  \nu\right)  -1}~\frac{r_{-}\left(
\nu\right)  }{\cos\frac{\pi j\left(  \nu\right)  }{2}}~.
\end{align*}

To conclude this section, we consider the contribution from a Regge pole
directly in the impact parameter representation. This can be easily achieved
by writing
\begin{equation}
\mathcal{B}\simeq-2\pi i\int d\nu~\beta\left(  \nu\right)  ~s^{j\left(
\nu\right)  -1}~\Omega_{i\nu}\left(  r\right)  ~.\label{eq10000}%
\end{equation}
It is then easy to show, using (\ref{eq3000})\ and (\ref{eq3100}), that the
contribution to the amplitude $A$ is given
\begin{align*}
A\left(  \mathbf{x},\mathbf{\bar{x}}\right)   & \simeq2\pi i~\mathcal{N}%
_{\Delta_{1}}\mathcal{N}_{\Delta_{2}}\int d\nu~\left(  -\right)  ^{j\left(
\nu\right)  }~4^{j\left(  \nu\right)  -1}~\Gamma\left(  \eta_{1}\left(
\nu\right)  \right)  \Gamma\left(  \eta_{2}\left(  \nu\right)  \right)
~\beta\left(  \nu\right)  \times\\
& \times~\int_{\mathrm{H}_{3}}d\mathbf{y}d\mathbf{\bar{y}~}\frac{\Omega_{i\nu
}\left(  \mathbf{y},\mathbf{\bar{y}}\right)  }{\left(  -2\mathbf{x\cdot
y}\right)  ^{\eta_{1}\left(  \nu\right)  }\left(  -2\mathbf{\bar{x}\cdot
\bar{y}}\right)  ^{\eta_{2}\left(  \nu\right)  }}~,
\end{align*}
where
\[
\eta_{i}\left(  \nu\right)  =2\Delta_{i}-1+j\left(  \nu\right)
~.\ \ \ \ \ \ \ \ \ \ \ \ \left(  i=1,2\right)
\]
In order to write the reduced amplitude $\hat{\mathcal{A}}=\left(
-\mathbf{x}^{2}\right)  ^{\Delta_{1}}\left(  -\mathbf{\bar{x}}^{2}\right)
^{\Delta_{2}}\,A$ in the Regge form (\ref{eq3200}), we must consider the
Fourier transform of the radial functions $\left(  -2\mathbf{x\cdot y}\right)
^{-\eta}$, which in general is given by the integral representation.
\begin{equation}
\frac{1}{\left(  -2\mathbf{x\cdot y}\right)  ^{\eta}}=\int d\nu~V\left(
\nu,\eta\right)  ~\Omega_{i\nu}\left(  \mathbf{x},\mathbf{y}\right)
~,\label{eq12345}%
\end{equation}
with%
\begin{align*}
V\left(  \nu,\eta\right)   & =\frac{4\pi}{\nu}\int_{0}^{\infty}d\rho~\sin
\nu\rho~\frac{\sinh\rho}{\left(  2\cosh\rho\right)  ^{\eta}}\\
& =\frac{\pi}{2}~\frac{\Gamma\left(  \frac{\eta-1+i\nu}{2}\right)
~\Gamma\left(  \frac{\eta-1-i\nu}{2}\right)  }{\Gamma\left(  \eta\right)  }~.
\end{align*}
It is then easy to show that the reduced amplitude is of the form
(\ref{eq3200}), with
\begin{equation}
\alpha\left(  \nu\right)  =V_{1}\left(  \nu\right)  ~\beta\left(  \nu\right)
~V_{2}\left(  \nu\right)  ~,\label{eq30000}%
\end{equation}
where we have defined the generalized couplings $V_{i}\left(  \nu\right)  $ as%
\begin{align*}
V_{i}\left(  \nu\right)   & =\mathcal{N}_{\Delta_{i}}~4^{j\left(  \nu\right)
-1}~\Gamma\left(  \eta_{1}\left(  \nu\right)  \right)  ~V\left(  \nu,\eta
_{i}\left(  \nu\right)  \right) \\
& =4^{j\left(  \nu\right)  -1}\frac{\Gamma\left(  \Delta_{i}-1+\frac{i}{2}%
\nu+\frac{1}{2}j\left(  \nu\right)  \right)  \Gamma\left(  \Delta_{i}%
-1-\frac{i}{2}\nu+\frac{1}{2}j\left(  \nu\right)  \right)  }{\Gamma\left(
\Delta_{i}\right)  \Gamma\left(  \Delta_{i}-1\right)  }~.
\end{align*}

\subsection{Spurious Poles\label{Sect100}}

In this section, we discuss the extra contributions in (\ref{eq1100}), hidden
in the omitted $\cdots$, which come from the poles of the partial waves
$\mathcal{T}_{E,J}$ in the physical region $\operatorname{Re}E\geq0$. These
contributions are subleading in the $\sigma\rightarrow0$ limit. Therefore, the
discussion which follows is included for completeness and can be skipped at a
first reading.

\begin{figure}
[ptb]
\begin{center}
\includegraphics[
height=1.1416in,
width=2.5344in
]%
{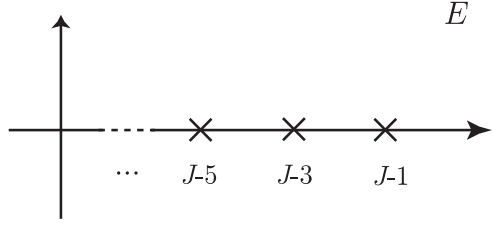}%
\caption{Spurious poles of $\mathcal{T}_{E,J}$ in the complex $E$--plane for
$\operatorname{Re}E>0$. They are located at $E=J-m$, with $m$ a positive odd
integer, and have residues proportional to $\mathcal{T}_{1+J,\,J-1-m}$.}%
\label{Fig3}%
\end{center}
\end{figure}

We first note that the chiral partial wave (\ref{eq2300}) has poles at%
\[
\operatorname*{Res}_{\bar{k}\rightarrow1-\frac{m}{2}}\mathcal{T}_{\bar{k}%
-1}=-\frac{1}{2\pi^{2}}\frac{\Gamma\left(  2-\bar{k}\right)  ^{4}}%
{\Gamma\left(  4-2\bar{k}\right)  \Gamma\left(  3-2\bar{k}\right)
}\mathcal{T}_{2-\bar{k}}~\ \ \ \ \ \ \ \left(  m=1,3,5,\cdots\right)
\]
Using the exact expression for $\mathcal{T}_{E,J}$ it is then immediate to
prove, as shown in figure \ref{Fig3}, that%
\begin{equation}
\operatorname*{Res}_{E\rightarrow1+J^{\prime}}\mathcal{T}_{E,J}=\frac{2}{\pi
}\,\tau\left(  J-J^{\prime}-1\right)  ~\mathcal{T}_{1+J,\,J^{\prime}%
}~,\label{eq5000}%
\end{equation}
where%
\[
J^{\prime}=J-2,\ J-4,~\cdots~\ .\ \ \ \ \ \ \ \ \left(  \operatorname{Re}%
J^{\prime}\geq-1\right)
\]%
Note that, at the residue point, the conformal Casimir for energy and spin
$\left(  E,J\right)  $ given by \cite{Osborn}
\[
E^{2}+\left(  J+1\right)  ^{2}-6
\]
is identical for the pairs $\left(  1+J^{\prime},J\right)  $ and $\left(
1+J,J^{\prime}\right)  $. Equation (\ref{eq5000}) then shows that the omitted
terms in (\ref{eq1100}) are given by%
\begin{equation}
-\frac{4}{\pi}~\sum_{J^{\prime}}~g_{J}\left(  -i\left(  1+J^{\prime}\right)
\right)  ~\left(  1+J^{\prime}\right)  ~\tau\left(  J+J^{\prime}+1\right)
\tau\left(  J-J^{\prime}-1\right)  ~\mathcal{T}_{1+J,J^{\prime}}%
~.\label{eq1200}%
\end{equation}
As anticipated in the previous section, the extra contributions have spins
$J^{\prime}<J$ and therefore do not effect the $\sigma\rightarrow0$ behavior
in (\ref{eq1400}). When one considers the full partial wave decomposition for
the reduced amplitude%
\[
\mathcal{A}=\sum_{J\geq0}\int d\nu~g_{J}\left(  \nu\right)  ~\mathcal{G}%
_{i\nu,J}~,
\]
one usually uses a representation in which the extra contributions from the
spurious poles in (\ref{eq1200}) are cancelled by contributions from poles of
$g_{J^{\prime}}\left(  \nu\right)  $. This is achieved by demanding that%
\[
4\, g_{J}\left(  -i\left(  1+J^{\prime}\right)  \right)  =-2\pi i~\frac
{1+J}{1+J^{\prime}}~\frac{\operatorname*{Res}g_{J^{\prime}}\left(  -i\left(
1+J\right)  \right)  }{\tau\left(  J-J^{\prime}-1\right)  }~~.~
\]
Let us note that the above relation is satisfied if
\[
g_J(\nu) = (k-\bar{k}+1)\frac{\Gamma(2k)}{\Gamma(k)^2}\frac{\Gamma(4-2\bar{k})}{\Gamma(2-\bar{k})^2}f(k,\bar{k})\ ,
\]
where 
\[ k+\bar{k}= 2+i\nu\ ,\ \ \ \ \ \ \ \ \ \ \ k-\bar{k}=J\ ,
\]
and where the function $f(k,\bar{k})$ satisfies, for $k-\bar{k}\in\mathbf{N}_0$,
\begin{eqnarray*}
&&f(k,\bar{k}) = f(2-\bar{k},2-k)\ ,\\
&&f(k,\bar{k}) = f(k,3-\bar{k})\ .\ \ \ \ \ \ \ \ \ \ \ \ \ \ \ \ \ \ \ \ \ \ (\bar{k}=5/2+\mathbf{N}_0)
\end{eqnarray*}

We conclude this section by discussing the role of these spurious poles in the
Sommerfeld--Watson resummation procedure. First we note that (\ref{eq5000})
immediately implies that
\begin{equation}
\operatorname*{Res}_{J\rightarrow\pm i\nu+m}\mathcal{G}_{i\nu,J}=\mp\frac
{i\nu}{\pi^{2}}\tau\left(  m\right)  \tau\left(  m\pm2i\nu\right)
~\mathcal{T}_{\pm i\nu+m+1,~\pm i\nu-1}\label{eq3900}%
\end{equation}
where $m$ is an odd positive integer. We then see that, when performing the
Sommerfeld--Watson transform, we get contributions from the poles in
(\ref{eq3900}), of the form%
\begin{align*}
& -\frac{1}{\pi}\sum_{m\in1+2\mathbb{N}_{0}}\int d\nu\frac{\nu}{\sinh\left(
\pi\nu\right)  }~\tau\left(  m\right)  \tau\left(  2i\nu+m\right)  ~~g_{\pm
}\left(  \nu,i\nu+m\right)  ~\times\\
& \times\left[  \mathcal{T}_{i\nu+m+1,~i\nu-1}\left(  \frac{z}{z-1},\frac
{\bar{z}}{\bar{z}-1}\right)  \pm\mathcal{T}_{i\nu+m+1,~i\nu-1}\left(
z,\bar{z}\right)  \right]  ~.
\end{align*}
We note that, although the above contributions come from poles at $J=\pm
i\nu+m$, the residues are given by partial waves of spin $-1\pm i\nu$, which
are subdominant compared with the leading Regge pole in the Lorentzian regime
in the $\sigma\rightarrow0$ limit. As shown in figure \ref{Fig1}, the above
contributions should therefore be considered as shadow contributions, and
should be added to the usual shadow contribution given by (\ref{eq3400}),
where the $J$ integral runs from $-1-i\infty$ to $-1+i\infty$.

\section{Multi--Reggeon Exchange in the Eikonal Limit\label{SecEikonal}}

In this section, we consider conformal field theories which admit a string
dual description. The amplitude $\mathcal{A}$ then has a perturbative
expansion in the string coupling%
\[
\mathcal{A}=1+\mathcal{A}_{\text{tree}}+\cdots~.
\]
In particular, we expect the tree level contribution $\mathcal{A}%
_{\text{tree}}$ to be dominated, in the Lorentzian regime, by an even Regge
pole at $J=j\left(  \nu\right)  $ corresponding to the graviton trajectory in
AdS, so that%
\[
\hat{\mathcal{A}}_{\text{tree}}\simeq2\pi i\int d\nu~\left(  -\right)
^{j\left(  \nu\right)  }~\alpha\left(  \nu\right)  ~\sigma^{1-j\left(
\nu\right)  }~\Omega_{i\nu}\left(  \rho\right)  ~.
\]
Using (\ref{eq20000}) and (\ref{eq10000}) we then quickly deduce that the
tree--level contribution to the phase shift $\Gamma$ is given by%
\begin{equation}
\Gamma\left(  s,r\right)  \simeq\int d\nu~\beta\left(  \nu\right)
~s^{j\left(  \nu\right)  -1}~\Omega_{i\nu}\left(  r\right)  ~,\label{eq40000}%
\end{equation}
where $\beta\left(  \nu\right)  $ is related to $\alpha\left(
\nu\right)  $ by the kinematical relation (\ref{eq30000}). As in
flat space \cite{ACV}, we expect that, for large impact parameter
$r$, the phase shift $\Gamma$ is dominated by the leading
contribution (\ref{eq40000}). In this limit, the
impact parameter representation%
\begin{align*}
A\left(  \mathbf{x},\mathbf{\bar{x}}\right)   & =\left(  -\right)
^{-\Delta_{1}}\left(  -\right)  ^{-\Delta_{2}}\mathcal{N}_{\Delta_{1}%
}\mathcal{N}_{\Delta_{2}}\times\\
& \times\int_{\mathrm{M}}\frac{d\mathbf{y}d\mathbf{\bar{y}}}{\left\vert
\mathbf{y}\right\vert ^{d-2\Delta_{1}}\left\vert \mathbf{\bar{y}}\right\vert
^{d-2\Delta_{2}}}~e^{-2i\mathbf{x\cdot y-}2i\mathbf{\bar{x}\cdot\bar{y}}%
}~e^{-2\pi i~\Gamma\left(  \mathbf{y},\mathbf{\bar{y}}\right)  }~
\end{align*}
resums multi--reggeon interactions in the eikonal limit, exactly as in flat.

The procedure we have followed to eikonalize reggeon exchanges
follows closely the analogous technique in flat space, restoring
unitarity at high energy. Our formul\ae \ therefore implicitly
incorporate unitarity requirements in AdS. The relation to
unitarity in the dual CFT has been partly explored in \cite{PS3,
HIM,AG}, but surely deserves a more thorough investigation.

\section{Example: $\mathcal{N}=4$ SYM at Large 't Hooft
Coupling\label{SecExample}}

In the remainder of the paper, we shall focus our attention on the canonical
example of $\mathrm{SU}\left(  N\right)  $, $\mathcal{N}=4$ super Yang--Mills
in $d=4$, which is dual \cite{Malda2} to Type IIB strings on \textrm{AdS}%
$_{5}\times\mathrm{S}^{5}$. Denoting with $\ell$ the \textrm{AdS} radius and
with $\sqrt{\alpha^{\prime}}$ the string length, the 't Hooft coupling%
\[
g^{2}\equiv g_{YM}^{2}N
\]
of the dual YM theory is given by%
\[
g=\frac{\ell^{2}}{\alpha^{\prime}}~.
\]
Moreover, the number of colors $N\,$\ is given, in terms of the \textrm{AdS}%
$_{5}$ gravitational constant $G$, by%
\[
\frac{1}{N^{2}}=\frac{2}{\pi}\frac{G}{\ell^{3}}~.
\]
One of the basic results of \cite{Paper3, PS2} is the fact that, in the
supergravity limit of large $g$, which neglects string corrections, the
leading eikonal gravitational interaction in AdS can be resummed. More
precisely, the phase shift $\Gamma\left(  \mathbf{y,\bar{y}}\right)  $, which
gives $\mathcal{B}=e^{-2\pi i\,\Gamma}$, is determined by the tree--level
interaction and reads%
\[
\Gamma=-\frac{2G}{\ell^{3}}\,s~\Pi_{2}\left(  r\right)  =-\frac{2G}{\ell^{3}%
}\,s\int\frac{d\nu}{\nu^{2}+4}~~\Omega_{i\nu}\left(  r\right)  ~,
\]
which corresponds to the exchange of the graviton dual to the dimension $4$
stress--energy tensor. As in flat space, in the full string theory this tree
level result is actually determined by the contribution from the full Regge
trajectory in the limit in which uniquely the graviton dominates. This
corresponds to a positive signature pole, and in general we shall have that%
\[
\Gamma=-\frac{2G}{\ell^{3}}\,\int d\nu~\beta\left(  \nu,g\right)  ~s^{j\left(
\nu,g\right)  -1}~\Omega_{i\nu}\left(  r\right)  ~,
\]
where we have shown the explicit dependence of $\beta$ and $j$ on the 't Hooft
coupling $g$. In particular, the Regge trajectory $j\left(  \nu,g\right)  $
determines the spectrum of lowest twist operators in the dual YM theory. More
precisely, in terms of the inverse $E\left(  j,g\right)  $ defined by
$j\left(  \pm iE\left(  j,g\right)  ,g\right)  =j$, the dimensions of the
lowest twist operators of increasing spin $J=2,4,6,\cdots$ is given by
$2+E\left(  J,g\right)  $. We recall some known facts about $j\left(
\nu,g\right)  $ in Appendix \ref{App}. In order to derive the leading string
corrections in \textrm{AdS} to $j$ and $\beta$, we consider the flat space
limit $\ell\rightarrow\infty$, where we know the result from the exact tree
level string computation in the Regge regime. More precisely, we must consider
the limit%
\begin{equation}
\nu=\ell q~,~\ \ \ \ \ \ \ \ \ \ \ \ \ \ \ g=\frac{\ell^{2}}{\alpha^{\prime}%
}~,~\ \ \ \ \ \ \ \ \ \ \ \ \ \ \left(  \ell\rightarrow\infty\right)
\label{eq4200}%
\end{equation}
with $q,\alpha^{\prime}$ fixed. Note that, in this limit, the variable $q$
measures momentum transfer, since $\ell^{-2}\left(  1+\nu^{2}\right)  $
measures the value of the Laplacian $-\ell^{-2}\square_{\mathrm{H}_{3}}$ on
the transverse space $\mathrm{H}_{3}\rightarrow\mathbb{E}^{3}$. Therefore,
matching the linear graviton Regge trajectory in flat space we must have that%
\begin{equation}
\lim_{\ell\rightarrow\infty}~j\left(  \ell q,\ell^{2}/\alpha^{\prime}\right)
=2-\frac{\alpha^{\prime}}{2}q^{2}~.\label{eq4000}%
\end{equation}
Moreover, we know that the graviton remains massless in \textrm{AdS} at any
value of the 't Hooft coupling $g$ due to the vanishing of the anomalous
dimension of the boundary stress energy tensor. Therefore we must also have
that%
\begin{equation}
j\left(  \pm2i,g\right)  =2~.\label{eq4100}%
\end{equation}
At strong coupling, we\ expand the general expression for the spin $j\left(
\nu,g\right)  $ in inverse powers of $g$ for small $\nu$ as%
\[
j\left(  \nu,g\right)  =\sum_{n\geq0}\frac{j_{n}\left(  \nu\right)  }{g^{n}%
}=\sum_{0\leq m\leq n}~j_{n,m}\frac{\nu^{2m}}{g^{n}},
\]
where we have assumed that $j\left(  \nu,g\right)  $ is regular at $\nu=0$. In
the above sum $m\leq n$ since the general term%
\[
\frac{\nu^{2m}}{g^{n}}=\mathcal{\ell}^{2\left(  m-n\right)  }\left(
\alpha^{\prime}\right)  ^{n}~q^{2m}%
\]
has a well--behaved $\ell\rightarrow\infty$ limit only in this case. Moreover,
the coefficients $j_{n,m}$ for $n=m$ are determined by the flat space limit
(\ref{eq4000}) to be $j_{0,0}=2$, $j_{1,1}=-1/2$ and $\ j_{n,n}=0 $ for
$n\geq2$. Equation (\ref{eq4100}) then implies that $j_{n}\left(
\pm2i\right)  =0$ for $n\geq1$. Since $j_{1}\left(  \nu\right)  =j_{1,0}%
+j_{1,1}\nu^{2}$, we then deduce that $j_{1,0}=-2$ and finally%
\begin{equation}
j\left(  \nu,g\right)  =2-\frac{4+\nu^{2}}{2g}-\left(  4+\nu^{2}\right)
\sum_{n\geq2}\frac{\tilde{j}_{n}\left(  \nu\right)  }{g^{n}}\label{eq4300}%
\end{equation}
with%
\[
~\tilde{j}_{n}\left(  \nu\right)  \text{ even polynomial of order }2n-4~.
\]
In particular, $\tilde{j}_{2}$ is just a constant and $j_{2}\propto j_{1}$.

A similar argument can be followed to deduce the leading string corrections to
$\beta\left(  \nu,g\right)  $. We have seen that, in the limit $g\rightarrow
\infty$ with $\nu$ fixed we must have that%
\[
\beta\sim\frac{1}{4+\nu^{2}}~.
\]
To recover the leading string corrections we must consider, on the other hand,
the limit (\ref{eq4200}) so that $-\pi\Gamma$ converges to the full phase
shift in flat space. In the limit we note that%
\begin{align*}
\lim_{\ell\rightarrow\infty}~\frac{1}{\ell^{2}}\Omega_{iq\ell}\left(  \frac
{R}{\ell}\right)   & =\frac{q}{4\pi^{2}R}\sin qR=\frac{q^{2}}{4\pi^{2}}%
j_{0}\left(  qR\right)  ~,\\
\lim_{\ell\rightarrow\infty}\frac{s}{\ell^{2}}  & =S~,
\end{align*}
where $R$ is the physical impact parameter in transverse Euclidean space
$\mathbb{E}^{3}$, $S$ is the physical Mandelstam energy--squared and
$q^{2}j_{0}\left(  qR\right)  $ computes Fourier transforms of radial
functions in $\mathbb{E}^{3}$. Then $\beta$ should converge to $\left(  \ell
q\right)  ^{-2}$ in the absence of string corrections, to reproduce the usual
gravitational interaction, and to the usual factor%
\[
\frac{\alpha^{\prime}}{4\ell^{2}}\left(  -\right)  ^{\frac{\alpha^{\prime}}%
{4}q^{2}}\frac{\Gamma\left(  \frac{\alpha^{\prime}}{4}q^{2}\right)  }%
{\Gamma\left(  1-\frac{\alpha^{\prime}}{4}q^{2}\right)  }%
\]
to reproduce the string corrections. Therefore we deduce that
\[
\beta\left(  \nu,g\right)  =-\frac{1}{4g}\left(  -\right)  ^{-j\left(
\nu,g\right)  /2}~\,\frac{\Gamma\left(  1-\frac{j\left(  \nu,g\right)  }%
{2}\right)  }{\Gamma\left(  \frac{j\left(  \nu,g\right)  }{2}\right)  }%
+\cdots~,
\]
with $\cdots$ vanishing in the limit (\ref{eq4200}).

We conclude by noting that the results derived in this section confirm and
extend the results derived in \cite{PS1}.

\section{Conclusions}

In this paper we have discussed the extension of Regge theory to
conformal field theories. Although inspired by the AdS/CFT
correspondence, most of the discussion is independent of the
existence of a gravity dual description of the CFT. On the other
hand, in the presence of a dual string formulation, we show how to
resum, in the eikonal limit of large impact parameter, multiple
reggeon interactions, following the analogous prescription used in
flat space. As an example, we have considered $\mathcal{N}=4$
$SU\left(  N\right)  $ supersymmetric Yang--Mills theory in $d=4$
at large 't Hooft coupling, where the gravi--reggeon pole
corresponds dually to the pomeron. We have in particular analyzed
the leading string corrections to pure graviton exchange. In a
subsequent work \cite{Paper5}, we plan to extend some of the
techniques of this paper to the weak coupling regime, connecting
to the formalism of BFKL \cite{BFKL} for hard pomeron exchange.

\section*{Acknowledgments}
We would like to thank  M.S. Costa, J. Penedones and G. Veneziano for discussions and comments.
Our research is supported in part by INFN, by the MIUR--COFIN contract 2003--023852, by the
MIUR contract \textquotedblleft Rientro dei cervelli\textquotedblright \ part VII.
We have been partially supported by the Galileo Institute
for Theoretical Physics, during the program \textit{String and M theory approaches to particle physics and
cosmology}, where part of this work was completed.

\vfill

\eject

\appendix

\section{Regge Theory in General Dimension $d$\label{AppBis}}

In this appendix, we wish to briefly outline the results of section
\ref{SecRegge} for general dimension $d$. The exchange of a primary of energy and spin
\begin{align*}
& \frac{d}{2}+E=k+\bar{k}~,\\
& J~=k-\bar{k}~,
\end{align*}
in the T--channel contributes to the reduced amplitude $\mathcal{A}$ as $
\mathcal{T}_{E,J}\,\left(  z,\bar{z}\right)  ~,$
were we normalize partial waves so that%
\[
\lim_{z\rightarrow0}\lim_{\bar{z}\rightarrow0}\mathcal{T}_{E,J}\sim\left(
-z\right)  ^{k}\left(  -\bar{z}\right)  ^{\bar{k}}~.
\]
The $\sigma\rightarrow0$ limit of the Lorentzian partial wave $\mathcal{\hat
{T}}_{E,J}$ is then given by%
\[
\mathcal{\hat{T}}_{E,J}\sim2\pi i~\left(  -\right)  ^{J}~\left(  \frac{\sigma
}{4}\right)  ^{1-J}~~\frac{\Pi_{E}\left(  \rho\right)  }{\tilde{\tau}\left(
E\right)  \tau\left(  \frac{d-4}{2}+J+E\right)  }~,
\]
where we have defined%
\[
\tilde{\tau}\left(  E\right)  =\pi^{\frac{4-d}{2}}\frac{\Gamma\left(
E+\frac{d-2}{2}\right)  }{\Gamma\left(  E+1\right)  }%
\]
and where%
\[
\Pi_{E}\left(  \rho\right)  =\frac{\tilde{\tau}\left(  E\right)  }{2\pi
}~e^{-\rho\left(  E+\frac{d-2}{2}\right)  }~F\left(  \frac{d}{2}%
-1,E+\frac{d-2}{2},E+1,e^{-2\rho}\right)
\]
is the scalar propagator in the transverse hyperbolic space $\mathrm{H}_{d-1}
$ of dimension $\frac{d-2}{2}+E$, canonically normalized. In the important
cases $d=2,4,6$ the propagator reads%
\begin{align*}
& \frac{1}{2}\frac{e^{-E\rho}}{E}%
~,\ \ \ \ \ \ \ \ \ \ \ \ \ \ \ \ \ \ \ \ \ \ \ \ \ \ \ \ \ \ \ \ \ \ \ \ \ \ \ \ \ \ \left(
d=2\right) \\
& \frac{1}{4\pi}\frac{e^{-E\rho}}{\sinh\rho}%
~,\ \ \ \ \ \ \ \ \ \ \ \ \ \ \ \ \ \ \ \ \ \ \ \ \ \ \ \ \ \ \ \ \ \ \ \ \ \ \ \ \left(
d=4\right) \\
& \frac{1}{8\pi^{2}}\frac{e^{-E\rho}\left(  \cosh\rho+E\sinh\rho\right)
}{\sinh^{3}\rho}~.\ \ \ \ \ \ \ \ \ \ \ \ \ \ \ \left(  d=6\right)
\end{align*}
Harmonic analysis of radial functions in $\mathrm{H}_{d-1}$ is performed using%
\begin{align*}
\Omega_{E}  & =\frac{E}{2\pi}\left(  \Pi_{E}-\Pi_{-E}\right)  \,\ \\
& =-\frac{1}{2^{d-1}\pi^{\frac{d+1}{2}}}E\sin\pi E\frac{\Gamma\left(
\frac{d-2}{2}+E\right)  \Gamma\left(  \frac{d-2}{2}-E\right)  }{\Gamma\left(
\frac{d-1}{2}\right)  }\times\\
& \times F\left(  \frac{d-2}{2}+E,\frac{d-2}{2}-E,\frac{d-1}{2},\frac
{1-\cosh\rho}{2}\right)  ~,
\end{align*}
which is given, in $d=2,4,6$, by%
\begin{align*}
& \frac{1}{2\pi}\cosh E\rho
~,\ \ \ \ \ \ \ \ \ \ \ \ \ \ \ \ \ \ \ \ \ \ \ \ \ \ \ \ \ \ \ \ \ \ \ \ \ \ \ \ \ \ \ \ \ \ \ \ \ \left(
d=2\right) \\
& -\frac{E}{4\pi^{2}}\frac{\sinh E\rho}{\sinh\rho}%
~,\ \ \ \ \ \ \ \ \ \ \ \ \ \ \ \ \ \ \ \ \ \ \ \ \ \ \ \ \ \ \ \ \ \ \ \ \ \ \ \ \ \ \ \ \left(
d=4\right) \\
& -\frac{E}{8\pi^{3}}\frac{\cosh\rho\sinh E\rho}{\sinh^{3}\rho}+\frac
{E^{2}\cosh E\rho}{8\pi^{3}\sinh^{2}\rho}%
~.\ \ \ \ \ \ \ \ \ \ \ \ \ \ \ \ \ \left(  d=6\right)
\end{align*}
General radial functions $G\left(  \mathbf{w},\mathbf{w}^{\prime}\right)  $
can then be expressed by%
\begin{align*}
G\left(  \mathbf{w},\mathbf{w}^{\prime}\right)   & =\int d\nu~g\left(
\nu\right)  ~\Omega_{i\nu}\left(  \mathbf{w},\mathbf{w}^{\prime}\right)  ~,\\
g\left(  \nu\right)   & =\frac{1}{\Omega_{i\nu}\left(  \mathbf{w}%
,\mathbf{w}\right)  }\int_{_{\mathrm{H}_{d-1}}}d\mathbf{w}^{\prime}%
~\Omega_{i\nu}\left(  \mathbf{w},\mathbf{w}^{\prime}\right)  ~G\left(
\mathbf{w}^{\prime},\mathbf{w}\right)  ~.
\end{align*}
Since, for general $d$, we have the limit for $\sigma\rightarrow0$%
\[
\tilde{\tau}\left(  E\right)  \tau\left(  {\textstyle \frac{d-4}{2}}+J+E\right)  ~\mathcal{T}%
_{E,J}~\ \ \ \ \ \ \ \rightarrow\ \ ~\ \ ~~~2\pi i~\left(  -\right)
^{J}\left(  \frac{\sigma}{4}\right)  ^{1-J}~~\Pi_{E}\left(  \rho\right)  ~,
\]
we can introduce the composite partial waves%
\[
\mathcal{G}_{E,J}=\frac{E}{2\pi}~\left[  \tilde{\tau}\left(  E\right)
\tau\left( {\textstyle \frac{d-4}{2}}+ J+E\right)  ~\mathcal{T}_{E,J}-\tilde{\tau}\left(  -E\right)
\tau\left({\textstyle \frac{d-4}{2}}+  J-E\right)  ~\mathcal{T}_{-E,J}\right]  ~,
\]
so that we have
\[
\mathcal{G}_{E,J}~\ \ \ \ \ \ \ \rightarrow\ \ ~\ \ ~~2\pi i~\left(  -\right)
^{J}\left(  \frac{\sigma}{4}\right)  ^{1-J}~~\Omega_{E}\left(  \rho\right)  ~.
\]
In particular, for $d=2$ we have that
\[
\mathcal{G}_{E,J}=\frac{1}{2}~\left[  
\tau\left(  J+E-1\right)  ~\mathcal{T}_{E,J}+
\tau\left(J-E-1\right)  ~\mathcal{T}_{-E,J}\right]\ .  \ \ \ \ \ \ \ \ \ \ \ \ \ \ \ (d=2)
\]

The full amplitude $\mathcal{A}$ will have then a decomposition in T--channel
partial waves given by%
\[
\mathcal{A}=\sum_{J\geq0}\int d\nu~g_{J}\left(  \nu\right)  ~\mathcal{G}%
_{i\nu,J}~.
\]

In the presence of a Regge pole, equations (\ref{eq3200}) and (\ref{eq10000})
are still valid, with the relation
between $\alpha\left(  \nu\right)  $ and $\beta\left(  \nu\right)  $ still
given by equation (\ref{eq30000}). The only equation which must be modified
involves $V\left(  \nu,\eta\right)  $, which has different expressions in
diverse dimensions. The function $V\left(  \nu,\eta\right)  $, still defined
by (\ref{eq12345}), is given in general by%
\[
V\left(  \nu,\eta\right)  =\frac{2\pi^{\frac{d-1}{2}}}{\Gamma\left(
\frac{d-1}{2}\right)  }\int d\rho~\sinh^{d-2}\rho~\frac{\Omega_{i\nu}\left(
\rho\right)  }{\Omega_{i\nu}\left(  0\right)  }~\left(  2\cosh\rho\right)
^{-\eta}~.
\]
The integral can be explicitly done and gives%
\[
V\left(  \nu,\eta\right)  =\frac{\pi^{\frac{d}{2}-1}}{2}~\frac{\Gamma\left(  \frac{\eta-\frac{d}{2}+1+i\nu}{2}\right)
~\Gamma\left(  \frac{\eta-\frac{d}{2}+1-i\nu}{2}\right)  }{\Gamma\left(  \eta\right)
}~.
\]
Therefore, the generalized couplings $V_i(\eta)$ in (\ref{eq30000}) are given in general by
\begin{align*}
V_{i}\left(  \nu\right)   & =\mathcal{N}_{\Delta_{i}}~4^{j\left(  \nu\right)
-1}~\Gamma\left(  \eta_{1}\left(  \nu\right)  \right)  ~V\left(  \nu,\eta
_{i}\left(  \nu\right)  \right) \\
& =4^{j\left(  \nu\right)  -1}\frac{\Gamma\left(  \Delta_{i}-\frac{d}{4}+\frac{i}{2}%
\nu+\frac{1}{2}j\left(  \nu\right)  \right)  \Gamma\left(  \Delta_{i}%
-\frac{d}{4}-\frac{i}{2}\nu+\frac{1}{2}j\left(  \nu\right)  \right)  }{\Gamma\left(
\Delta_{i}\right)  \Gamma\left(  \Delta_{i}+1-\frac{d}{2}\right)  }~.
\end{align*}

\section{Known Facts About $j\left(  \nu,g\right)  $\label{App}}

In this appendix, we recall some basic known facts about the Regge trajectory
$j\left(  \nu,g\right)  $ which determines the spectrum of lowest twist
operators in the dual YM theory. The strong and weak coupling limit of $j$ are
given respectively by
\begin{align*}
\lim_{g\rightarrow\infty}~j\left(  \pm iE,g\right)   & =2~,\\
\lim_{g\rightarrow0}~j\left(  \pm iE,g\right)   & =\max\left(  \left\vert
E\right\vert ,1\right)  ~.
\end{align*}%
\begin{figure}
[ptb]
\begin{center}
\includegraphics[
height=1.9069in,
width=3.6752in
]%
{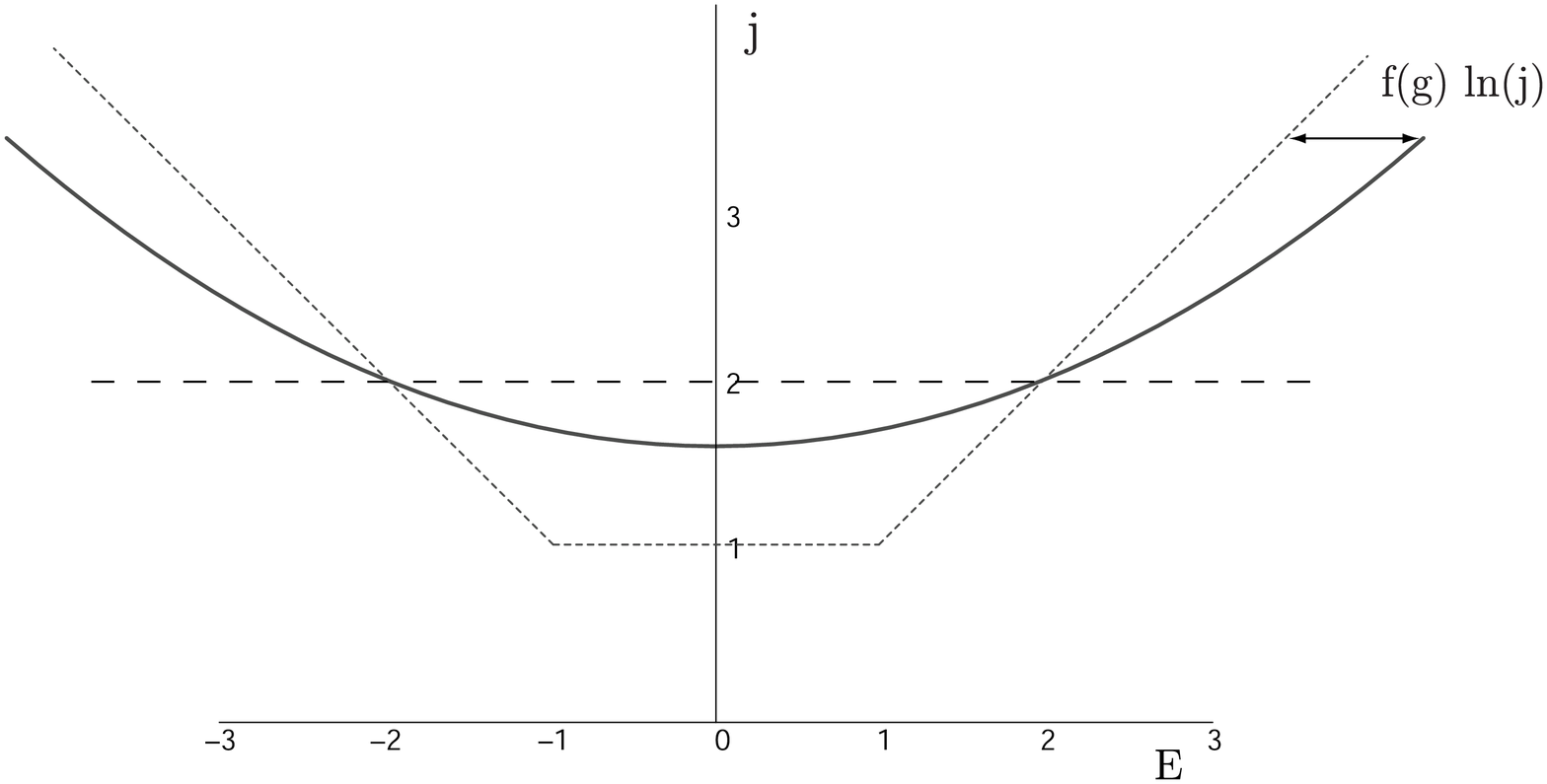}%
\caption{Shown is the curve determining the spectrum of lowest twist operators
in $\mathcal{N}=4$ SYM at large $N$. The $g\rightarrow0$ and $g\rightarrow
\infty$ limits are shown with fine and normal dash, whereas the curve for a
finite generic value of the 't Hooft coupling $g$ is shown with a continuous
line. The leading behavior $f\left(  g\right)  \ln j$ of the anomalous
dimension for large $j$ is also shown. }%
\label{Fig5}%
\end{center}
\end{figure}
The inverse function $E\left(  j,g\right)  $ has a universal large $j$
behavior%
\[
\lim_{j\rightarrow\infty}~E\left(  j,g\right)  =j+f\left(  g\right)  \ln
j+\dots
\]
determined by the so--called cusp anomalous dimension $f\left(  g\right)  $.
The function $f\left(  g\right)  $ is known to all orders \cite{BES}, with
well--defined strong and weak coupling expansions
\begin{align*}
f\left(  g\right)   & =\frac{g}{\pi}-\frac{3\ln2}{\pi}+\cdots~,\\
f\left(  g\right)   & =\frac{1}{2\pi^{2}}\left(  g^{2}-\frac{1}{48}g^{4}%
+\frac{11}{11520}g^{6}+\cdots\right)  ~.
\end{align*}
Finally the functions $E\left(  j,g\right)  $ and $j\left(  \pm iE,g\right)  $
have weak coupling expansion given by
\begin{align*}
E\left(  j,g\right)   & =j+\frac{g^{2}}{2\pi^{2}}\left(  \Psi\left(
j-1\right)  -\Psi\left(  1\right)  \right)  +\mathcal{\cdots}~,\\
j\left(  \pm iE,g\right)   & =1+\frac{g^{2}}{4\pi^{2}}\left(  2\Psi\left(
1\right)  -\Psi\left(  \frac{1+E}{2}\right)  -\Psi\left(  \frac{1-E}%
{2}\right)  \right)  +\cdots~,
\end{align*}
where the second result is the celebrated BFKL\ spin \cite{BFKL} for the hard
perturbative QCD pomeron. Some of this information is graphically summarized
in figure \ref{Fig5}, following \cite{PS1}.


\end{document}